\documentclass{ws-procs9x6}

\def\be{\begin{equation}}
\def\ee{\end{equation}}
\def\ba{\begin{eqnarray}}
\def\ea{\end{eqnarray}}

\def\sfrac#1#2{{\textstyle \frac{#1}{#2}}}

\begin{document}

\title{
A COVARIANT FORMALISM FOR THE $N^\ast$
ELECTROPRODUCTION AT HIGH MOMENTUM TRANSFER}

\author{G. RAMALHO$^\dagger$}

\address{Centro de F{\'\i}sica Te\'orica de Part{\'\i}culas,
Av.\ Rovisco Pais, 1049-001 Lisboa, Portugal \\
$^\dagger$E-mail: gilberto.ramalho@cftp.ist.utl.pt}

\author{F. GROSS}

\address{Thomas Jefferson National Accelerator Facility, Newport News,
VA  23606, USA  \\
College of William and Mary, Williamsburg, Virginia 23185, USA}

\author{M. T. PE\~NA}

\address{Centro de F{\'\i}sica Te\'orica de Part{\'\i}culas,
Av.\ Rovisco Pais, 1049-001 Lisboa, Portugal \\
Departamento de F{\'\i}sica,
Instituto Superior T\'ecnico, 1049-001 Lisboa, Portugal}

\author{K. TSUSHIMA}

\address{EBAC in Theory Center,
Thomas Jefferson National Accelerator Facility, \\ Newport News,
Virginia  23606, USA}

\begin{abstract}
A constituent quark model based
on the spectator formalism is applied to the
$\gamma N \to N^\ast$ transition for the three cases, where $N^\ast$ is
the nucleon, the $\Delta$ and the Roper resonance.
The model is covariant, and therefore can be used for
the predictions
at higher four-momentum transfer squared, $Q^2$.
The baryons are described as an off-mass-shell
quark and a spectator on-mass-shell diquark systems.
The quark electromagnetic current is described by
quark form factors, which have a form inspired by
the vector meson dominance.
The valence quark
contributions of the model are
calibrated by lattice QCD simulations and experimental data.
Contributions of the meson cloud to
the inelastic processes are explicitly included.
\end{abstract}


\bodymatter

\section{Introduction}\label{secIntroduction}

Study of the nucleon structure and
its electromagnetic excitation
is one of the important topics associated with
the missions and activities of modern accelerator facilities.
At Jefferson lab
very accurate data have been extracted for the 
$\gamma N \to N^\ast$ reactions, for several $N^\ast$ resonances
at low and high $Q^2$ [\refcite{Burkert04,CLAS}],
defining new challenges for the theoretical models.
Although one believes that the nucleon excitations
are governed
by QCD with quarks and gluons
in a non-perturbative regime,
it is at present nearly impossible to solve QCD exactly in the
region $Q^2=0-10$ GeV$^2$. Thus,
one has to rely on some effective and phenomenological approaches.
One of popular approaches is the   
dynamical coupled channel reaction models
[\refcite{Diaz07b,Kamalov01,Julich,Bonn}], 
where the effective degrees of freedom
are mesons and baryons.
In these models a baryon core structure 
is assumed, and it is modified by the meson cloud dressing
resulting from the meson-baryon interactions.
Effective field theories based on chiral
symmetry, with pions and baryons alone as
degrees of freedom, are applicable only
in the very low $Q^2$ region.
On the other hand,
perturbative QCD works only in the 
very large $Q^2$ region [\refcite{NDelta,NDeltaD}].
Alternative descriptions are constituent quark models
[\refcite{Giannini91}].
A constituent quark has an internal structure
resulting from the quark-antiquark
dressing, and from the short range interaction with gluons.
The quark structure of a baryon   
can be represented by electromagnetic valence quark form factors.
In this work we present the covariant
spectator quark model
[\refcite{Nucleon,NDelta,NDeltaD}],
and show several applications of the model.
Covariance is important in
the applications in the higher $Q^2$ region.

\begin{figure}[t]
\centerline{
\mbox{
\includegraphics[width=6.0cm]{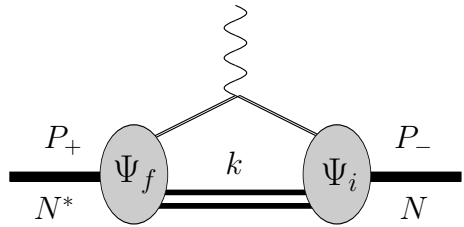} }}
%\vspace{-1cm}
\caption{$\gamma N \to N^\ast$ transition in the
covariant spectator quark model (diquark on-shell) in
relativistic impulse approximation.
$P_+$ ($P_-$) represents the final (initial)
baryon momentum and $k$ the intermediate diquark momentum.
The baryon wave functions
are represented by $\Psi_f$ and $\Psi_i$
for the final and initial states, respectively.}
\label{figImp}
\end{figure}

\section{Spectator quark model}

In the covariant spectator quark model
baryons are described as a three-valence
quark systems with an on-shell quark-pair,
or diquark, while the remaining quark is off-shell
and free to interact with electromagnetic fields.
The quark-diquark vertex is then represented
by a baryon $B$ wave function $\Psi_B$ that
effectively describes quark confinement [\refcite{Nucleon}].
See Fig.~\ref{figImp}.
%

% REMOVE TITLES

%\subsection{Quark current}

The 
%photon-quark interaction 
quark electromagnetic current
$j_I^\mu$ is given
by the Dirac and Pauli structures:
\be
j_I^\mu=
\left(\frac{1}{6}f_{1+} +
  \frac{1}{2}f_{1-} \tau_3 \right)
\left(
\gamma^\mu -\frac{\not \! q q^\mu}{q^2}  \right)
+ 
\left(
\frac{1}{6}f_{2+} +
  \frac{1}{2}f_{2-} \tau_3
\right) \frac{i \sigma^{\mu \nu} q_\nu}{2M_N},
\label{eqJI}
\ee
where $M_N$ is the nucleon mass,
$f_{1\pm}$ and $f_{2\pm}$ are the quark form
factors as functions of $Q^2$, and $\tau_3$ the isospin operator.
To represent the quark structure 
%one considers a vector meson dominance parametrization,
we adopt a vector meson dominance motivated parametrization,
where the form factors are written in terms of two
vector meson poles:
\ba
%\hspace{-.8cm}
& &
f_{1\pm}(Q^2)=  \lambda_q + (1-\lambda_q)
\frac{m_v^2}{m_v^2+ Q^2} +
c_{\pm}\frac{Q^2 M_h^2}{\left(M_h^2+Q^2\right)^2}
\label{eqF1}
\\
%\hspace{-.8cm}
& &
f_{2\pm}(Q^2)= \kappa_\pm
\left\{
d_\pm \frac{m_v^2}{m_v^2+ Q^2} +
(1-d_{\pm}) \frac{Q^2 }{M_h^2+Q^2} \right\}.
\label{eqF2}
\ea
In the above $m_v = m_\rho$ is a light vector meson mass
that effectively represents the $\rho$ and $\omega$ poles
and $M_h$ is the  
an effective heavy vector meson mass,
that takes into account the short range phenomenology.
We chose $M_h= 2 M_N$ in the present study.
The isoscalar $\kappa_+$ and isovector $\kappa_-$
quark anomalous moments are fixed by the
nucleon magnetic moments.
The adjustable parameters are $\lambda_q$
and the mixture coefficients $c_\pm$ and $d_\pm$.
In the study of the nucleon properties, 
it turned out that $d_+=d_-$ gives a very
good description of the nucleon
electromagnetic form factor [\refcite{Nucleon}].
This reduces the number of adjustable parameters to 4.
The quality of the  
model description for the nucleon form factors
is illustrated
in Fig.~\ref{figNucleon}.
The quark current fixed by the nucleon form factors
will be used for all other applications discussed below.

% Figure 2

% REMOVE TITLES

%\subsection{Baryon wave functions}

To write the  baryon $B$ wave function $\Psi_B$,
we start 
from the baryon rest frame, 
$P=(M_B,0,0,0)$, with $M_B$ the baryon mass.
We represent the baryon wave function
as the direct product of the
diquark and quark states of flavor, spin, 
orbital angular momentum and radial excitation, 
consistent with the baryon quantum numbers.
The flavor  
states are written
using the SU$_F$(3) quark states $\Phi_I^{0,1}$,  with the diquark
of total isospin $I=0,1$.
Similarly, the diquark spin states
associated with spin $S=0,1$, $\Phi_S^{0,1}$,
can be written in terms of the
polarization vectors
$\varepsilon^\mu(0)=(1,0,0,0)$
and $\varepsilon^\mu(\pm 1)=\mp \sfrac{1}{\sqrt{2}}(0,1,\pm i,0)$,
where $\lambda= 0,\pm 1$ is the diquark polarization
[\refcite{Nucleon,NDelta,FixedAxis}].
Once the wave functions are written explicitly
in terms of the baryon properties in the rest frame,
the relativistic generalization is
performed with a boost to the moving frame.
The diquark polarization vectors  will 
be represented by a function $\varepsilon_P^\mu(\lambda)$ of the center-of-mass
momentum $P$ in the fixed-axis representation, as
described in Ref.~[\refcite{FixedAxis}].
The explicit covariant form for the nucleon,
$\Delta$ and Roper wave functions can be found
in Refs.~[\refcite{Nucleon,NDelta,NDeltaD,Roper}].

% REMOVE TITLES

%\subsection{Electromagnetic transition current}

The electromagnetic current associated with
the final state $N^\ast$  in the covariant spectator
quark model (see  Fig.~\ref{figImp}) is determined by
\be
J^\mu=
3 \sum_\lambda \int_k
\overline \Psi_f (P_+,k) j_I^\mu \Psi_i(P_-,k).
\label{eqJ}
\ee
In the  
above, $\int_k$ represents
the covariant integral  
with respect to the on-mass-shell diquark momentum 
and $\lambda$ the diquark polarization.
For simplicity,
diquark and baryon polarization indices are
suppressed.

\section{Applications}

In Eq.~(\ref{eqJ}) 
we can write the electromagnetic transition current in
terms of $q=P_+-P_-$ and $P=\sfrac{1}{2}(P_+ + P_-)$. 
The corresponding form factors, invariant functions of $Q^2$, are
$G_E$ and $G_M$ for the nucleon, 
$G_M^\ast$, $G_E^\ast$ and $G_C^\ast$ for the $\Delta$, 
and $F_1^\ast$ and $F_2^\ast$ for the Roper.

\subsection{Nucleon}

\begin{figure}[t]
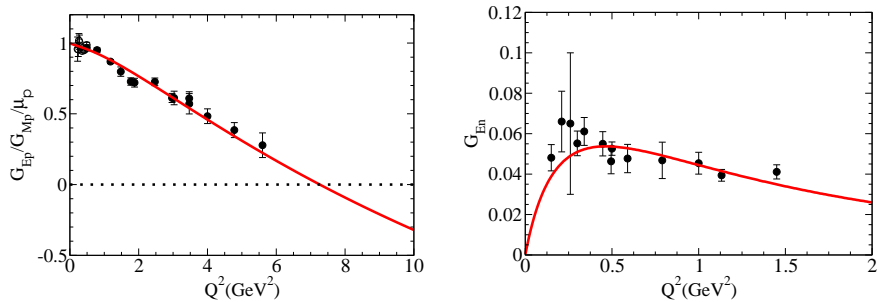

\centerline{
\mbox{
\includegraphics[width=5.5cm]{GEp.eps}
\hspace{.3cm}
\includegraphics[width=5.5cm]{GEn.eps} }}
\caption{Ratio for the electric and magnetic
form factors for the proton, and neutron electric form factor
for model II in Ref.~[\refcite{Nucleon}].
Data are as presented in Ref.~[\refcite{Nucleon}].}
\label{figNucleon}
\end{figure}

For the nucleon, the simplest wave function
%corresponds to 
has a quark-diquark S-wave configuration
[\refcite{Nucleon}]:
\be
\Psi_N= \frac{1}{\sqrt{2}}
\left[\Phi_I^0 \Phi_S^0 + \Phi_I^1 \Phi_S^1
\right] \psi_N(P,k),
\label{eqNucleon}
\ee
with $\Phi_I^{0,1}$ and  $\Phi_S^{0,1}$,  
the diquark spin and isospin states 
of 0 and 1, and $\psi_N$  a scalar
wave function.
Results for the nucleon form factors ~[\refcite{Nucleon}]
are shown in Fig.~\ref{figNucleon}.
No explicit pion cloud  is included for the results. 

% Figure 2 (previous position)

\subsection{$\gamma N \to \Delta$ transition}

The $\gamma N \to \Delta$ transition is more complex
than the nucleon elastic reaction.
The transition current (\ref{eqJ}), with  $\Psi_\Delta$,
associated exclusively with the quark valence degrees of freedom, 
is insufficient to explain the data [\refcite{NDelta,NDeltaD}]. 
As near the $\Delta$ region the 
nucleon has enough energy to create a pion,
the electromagnetic interaction with intermediate 
pion-baryon states should also be considered. 
Then, the transition form factors can 
be decomposed as
\be
G_X^\ast = G_X^b + G_X^\pi,
\ee
where $G_X^b$ stands for the contribution 
of the quark core (bare) and 
$G_X^\pi$ for the contribution due to the pion cloud.
The label $X$ holds for $M$ (magnetic dipole),
$E$ (electric quadrupole) and
$C$ (Coulomb quadrupole) form factors.
This decomposition is justified when the pion
is created by the overall baryon three-quark system
and not from a single quark.

\begin{figure}[t]
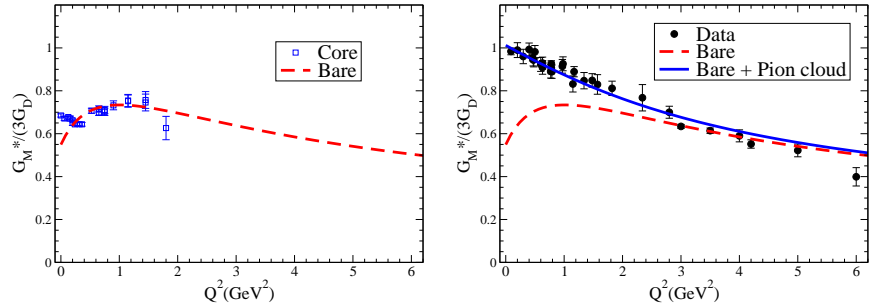

\vspace{.2cm}
%\vspace{-2cm}
\centerline{
\mbox{
\includegraphics[width=5.5cm]{GM2a.eps}\hspace{.3cm}
\includegraphics[width=5.5cm]{GM2b.eps}
}}
\caption{$G_M^\ast$ normalized by $3 G_D$ in
the $\gamma N \to \Delta$ transition.
Left: Valence quark contributions 
from our model compared with
the bare contributions of Sato-Lee model
[\refcite{Diaz07b}].
Right: Bare plus phenomenological pion cloud contributions, 
compared with the data
[\refcite{NDelta}].}
%\vspace{-2cm}
\label{figGM}
\end{figure}

As a first application we 
describe the $\Delta$  as a quark-diquark  S-state
coupled to a spin $3/2$  to
form a total $J=3/2$ state [\refcite{NDelta}].
The transition proceeds  
then only through the magnetic form factor [\refcite{NDelta,NDeltaD}]:
\be
G_M^b(Q^2)= 4 \eta f_v {\cal I},
\label{eqGMb}
\ee
where $\eta= \sfrac{2}{3\sqrt{3}}\frac{M_N}{M_N+ M_\Delta}$,
$f_v= f_{1-}+ \sfrac{M_N+ M_\Delta}{2 M_N} f_{2-}$
and ${\cal I}$ is the overlap integral between
the nucleon and $\Delta$ S-state scalar wave functions.
This result allows us to understand why the
pion cloud is essential to describe the data,
and necessary to be added.
In a pure constituent quark model the overlap integral
is limited by the wave function normalization  [\refcite{NDelta}].
At $Q^2=0$, ${\cal I} \le 1$,
and for the spectator quark model this implies
an upper value for $G_M^b(0)$ of 2.07,  to be compared with the
experimental result 3.02  [\refcite{NDelta}].
Higher
angular momentum partial waves
for the relative quark-diquark motion are only possible to contribute
to the quadrupole form factors.
Since these are small compared to $G_M^*$,
they have a reduced weight in 
the wave function, and
consequently in $G_M^*$. Therefore,
the discrepancy found in the leading form factor, 
between constituent quark models and experimental data,
is mainly to be 
compensated by the pion cloud contributions.
To adjust the 
valence quark contributions  
we use the results of the Sato-Lee model
obtained from the data [\refcite{Diaz07b}], 
subtracted by the pion cloud contributions.
The result of the fit is presented in the left panel of Fig.~\ref{figGM}.
The experimental data points are  
reached when
$G_M^\pi=\lambda_\pi \left(\sfrac{\lambda_\pi^2}{\Lambda_\pi^2 + Q^2}\right)^2
(3 G_D)$ is added to $G_M^b$ 
($G_D$ the nucleon dipole factor).
See the right panel in Fig.~\ref{figGM}.

% Figure 3 (previous position)

\begin{figure}[t]
\centerline{
\mbox{
\includegraphics[width=5.5cm]{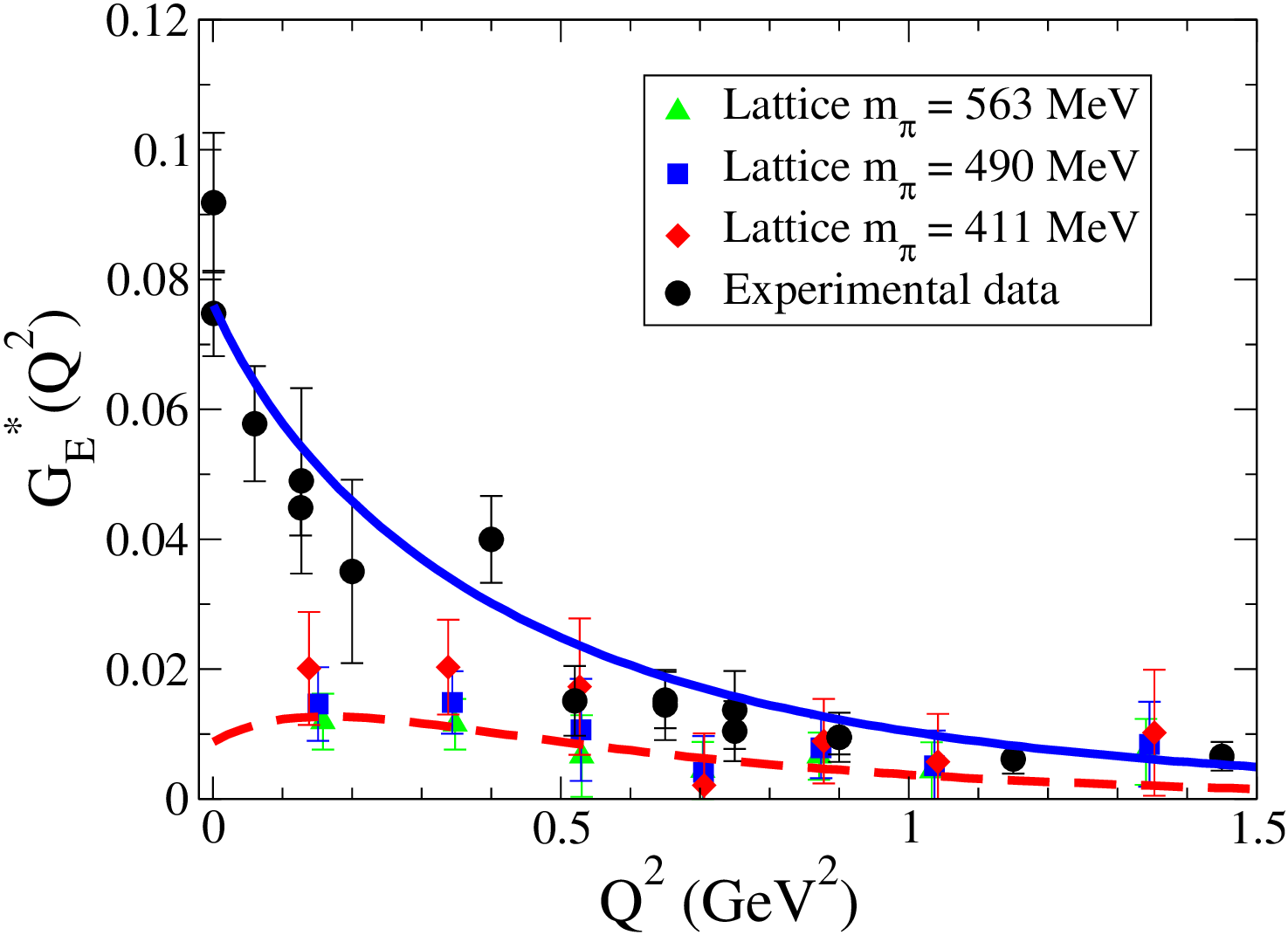}
\hspace{.3cm}
\includegraphics[width=5.5cm]{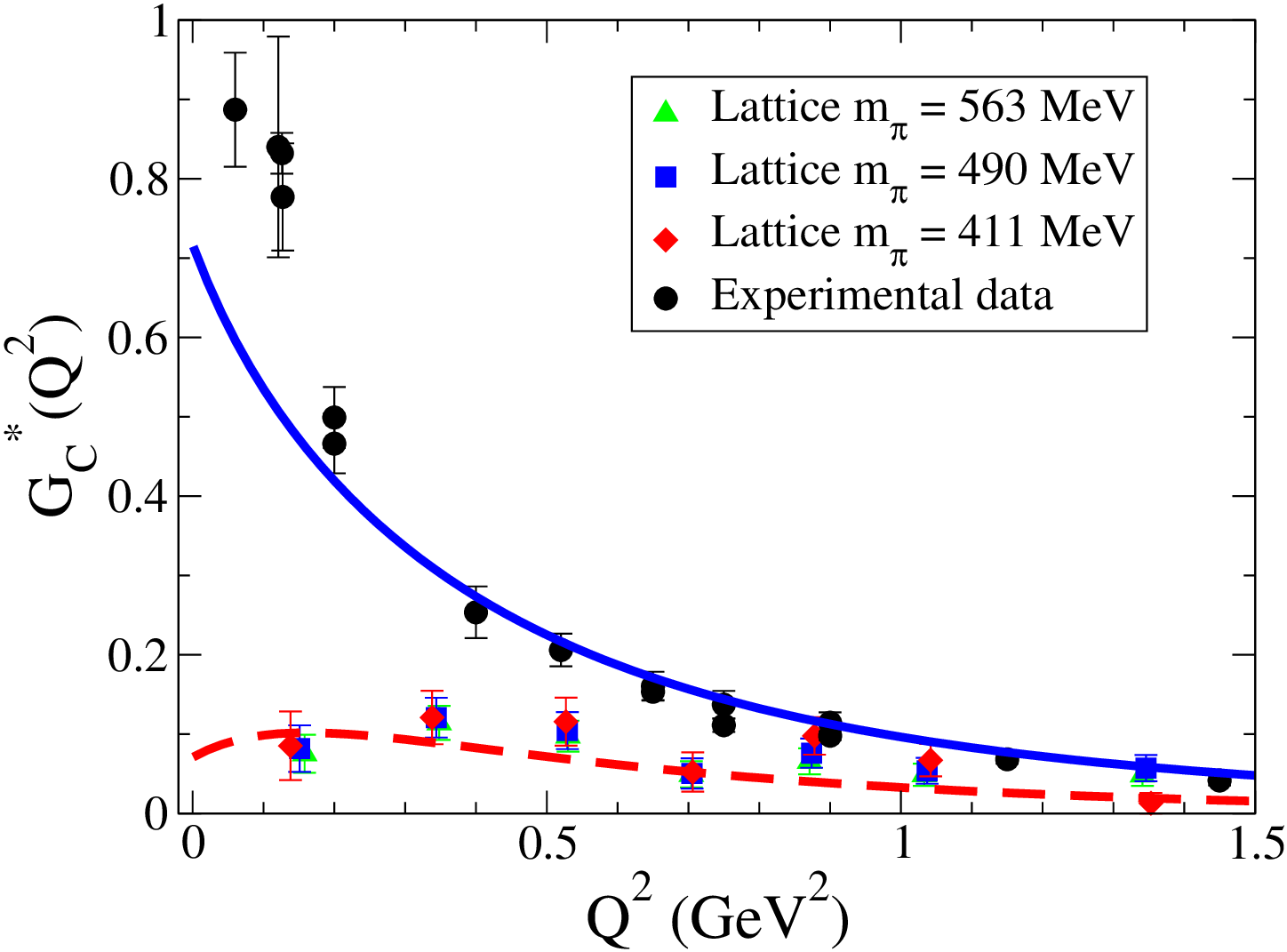}}}
\caption{
Electric and Coulomb quadrupole form factor [\refcite{LatticeD}] 
for the $\gamma N \to \Delta$ transition.
Lattice data are taken from [\refcite{Alexandrou08}].}
\label{figGEGC}
\end{figure}

The next step is to include D-state admixtures in the wave function
\mbox{
as [\refcite{NDeltaD}],} 
\be
\Psi_\Delta= N \left[
\Psi_S + a \Psi_{D3} + b \Psi_{D1}
\right],
\ee
where $\Psi_{D3}$ represents a D-state with a core spin 1/2 
and $\Psi_{D1}$ a D-state with a core spin 3/2. %[\refcite{NDeltaD}].
The D-state generates contributions for $G_E^\ast$ and
$G_C^\ast$ form factors, which, otherwise for a pure S-wave function would
vanish identically. To separate the valence quark contributions, 
we have also extended the model to the lattice QCD regime
[\refcite{Lattice,LatticeD,Omega}]
and adjusted the D-state parameters to the
quenched lattice QCD data [\refcite{Alexandrou08}] 
for a pion mass region where pion cloud effects are
expected to be small [\refcite{LatticeD}].
Once the valence quark contributions are fixed from the lattice regime, the
results are extrapolated back to the physical region.
Finally, by adding the pion cloud contributions derived from the
large-$N_c$ limit
[\refcite{NDeltaD,LatticeD}] to the valence quark contributions $G_X^b$, 
we obtain the final result
shown in Fig.~\ref{figGEGC}. The results agree well with the physical data.
See Refs.~[\refcite{NDeltaD,LatticeD}] for details.

% Figure 4 (previous position)

\subsection{$\gamma N \to$ Roper transition}

\begin{figure}[t]
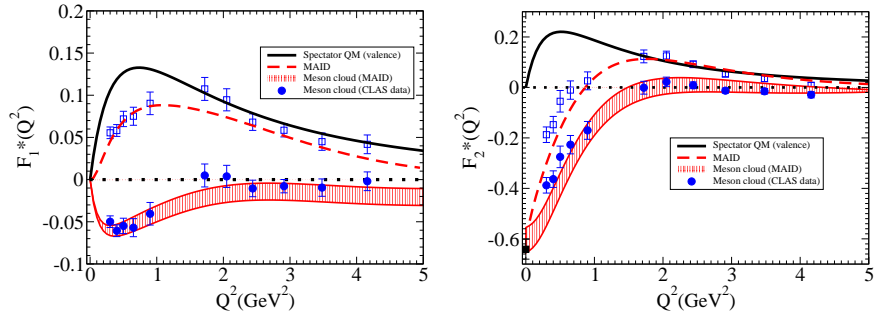

\vspace{.2cm}
\centerline{
\mbox{
\includegraphics[width=2.2in]{F1subMc}  \hspace{.1cm}
\includegraphics[width=2.2in]{F2subMc} }}
\caption{
$\gamma N \to$ Roper transition form factors, $F_1^\ast$ and $F_2^\ast$.
Data are from [\refcite{CLAS}].
Meson cloud contributions based
on the MAID fit [\refcite{MAID}] are represented by the band 
[\refcite{Roper}].
The meson cloud contributions from CLAS
are also shown.}
\label{figRoper}
\end{figure}

Within the covariant spectator quark model,
we can also describe the Roper system,
as the first radial excitations of the nucleon [\refcite{Roper}].
Thus, the Roper wave function has
the same structure as that for the nucleon Eq.~(\ref{eqNucleon}),
except for the scalar wave function, which is replaced by $\psi_R$.
Under this assumption, the orthogonality
between the Roper and nucleon wave functions
is reduced to the orthogonality
between the corresponding scalar wave functions:
$\int_k \psi_R \psi_N=0$ at $Q^2=0$.
This fixes the free parameters
in $\psi_R$ completely, assuming that the nucleon
and the Roper have the same short range behavior, 
but differ in the long range structure.
No extra parameter is needed additionally to
the ones already fixed in the
nucleon wave function [\refcite{Roper}].
Once $\psi_R$ is defined,  we can calculate and predict 
the nucleon to Roper transition form factors $F_1^\ast$ and $F_2^\ast$.
The results are shown in Fig.~\ref{figRoper}, and are
consistent with the CLAS data [\refcite{CLAS}]
for $Q^2 > 2$ GeV$^2$.
These facts support the idea that the valence
quark degrees of freedom are well described
in the covariant spectator quark model.
Once the valence quark contributions are determined,
we can then  estimate the meson cloud contributions  
using the decomposition
$F_i^\ast= F_i^b + F_i^{mc} (i=1,2)$,
where $F_i^b$ is the bare contribution
and $F_i^{mc}$ is the meson cloud contribution [\refcite{Roper}].
The results are also in Fig.~\ref{figRoper}.
%(See Fig.~\ref{figRoper}).
%The meson cloud contributions from the CLAS data
%and MAID parametrization are showed in Fig.~\ref{figRoper}.

\section{Conclusions}

We have developed a formalism which is successful in
describing the valence quark contributions to the
nucleon form factors, without the inclusion of pion cloud.
The present approach also describes very well 
the $\gamma N \to \Delta$ data,
both in the physical regime and
the lattice regime, where the pion cloud effects 
are suppressed in the lattice regime.
Furthermore, the results are consistent with the
estimate of the core contributions of the Sato-Lee model.
As for the  
$\gamma N \to \mbox{Roper}$ transition,  
we have obtained a very good
description for the high $Q^2$ data,
where valence quark degrees of freedom are
expected to be dominant.

Other applications of  
the present approach 
have been made also for the determination of the
$\Delta$ [\refcite{DeltaFF,DeltaDFF0,DeltaDFF}]
and decuplet [\refcite{Omega}] electromagnetic form factors,
and the octet magnetic moments 
(in this case including pion cloud effects).

\vspace{0.5cm}
\noindent
{\bf Acknowledgments:}

G.~R.\ would like to thank Viktor Mokeev for the invitation to the workshop.
G.~R.\ was supported by the Portuguese Funda\c{c}\~ao para
a Ci\^encia e Tecnologia (FCT) under the grant
SFRH/BPD/26886/2006.

%\vspace{-.5cm}

\bibliographystyle{ws-procs9x6}
\bibliography{biblo}

\end{document}